\documentstyle[prd,aps,epsfig]{revtex}

\begin{document}
\draft

\title{Quark combinatorics  for production ratios in hadronic
$Z^0$ decays}
\author{V. A. Nikonov}
\address{Petersburg Nuclear Physics Institute, Gatchina, St.
Petersburg 188350, Russia}
\author{and J. Nyiri}
\address{Central Research Institute for Physics, Budapest, Hungary}
\date{June 19, 2000}
\maketitle

\begin{abstract}
Model independent verification of the
quark combinatorics rules,
which govern the ratios of the yields of secondaries,
is presented for jet processes. Because of a large number of
produced resonances in the hadron jets, a test of the
quark combinatorics rules is hardly possible in the central
region, $x_{hadron} \leq 0.2$. However, a
model-independent verification is plausible at $x_{hadron} \sim 1$.
It is shown that for the large-$x_{hadron}$
kinematical region the quark combinatorial
relations are in a reasonable agreement with data for
$\rho^0/\pi^0$ and $p/\pi^+$ ratios.
\end{abstract}

\vspace{2cm}

Initially the rules of quark combinatorics were suggested in \cite{1,2}
and checked in multiple production processes, see \cite{3} and
references therein. Henceforth these rules were applied to the
decay processes of resonances \cite{4,5,6,7}.

Concerning  multiple hadron
production processes, it is necessary to underline
that quark combinatorics rules manifest important features of the
process in which quarks form hadrons,
that is, soft colour neutralization and soft
hadronization \cite{3}.

There exist two types of predictions of quark combinatorics for
multiple production processes.
The first one is the prediction of the
yield ratios for hadrons belonging to the same quark multiplet. For
example, we have for the lightest meson nonet:
\begin{equation}
\rho^0/\pi^0=3, \;\;\; K^{*0}/K^0=3, \;\;\; K^{*\pm}/K^\pm=3.
\end{equation}
Two statements form the basis for these relations: $(i)$ wave functions
of hadrons from the same multiplet are equal, and  $(ii)$ quarks which
are fused into hadrons are spin- and flavour-non-correlated.

The second type of prediction is related to the yields of mesons and
baryons. For the hadronization of the quark $q_i$ in $(q,\bar q)$-sea
the production rule reads:
\begin{equation}
q_i+(q,\bar q)_{\mbox {sea}} \to \frac1{3}B_i+\frac{2}{3}M_i+
\frac1{3}M+(M,B,\bar B)_{\mbox {sea}},
\end{equation}
where $B_i$ and $M_i$ are baryons and mesons containing the quark $q_i$
and $M$, $B$ and $\bar B$ are mesons and baryons of the sea.

The prompt verification of Eqs. (1) and (2) is rather difficult because
in multiple production processes a number of resonances is produced.
Generally one can write
\begin{equation}
M\ =\ \sum_L\mu_L M_L \quad \mbox{ and }\quad B\ =\
\sum_L\beta_LM_L\ ,
\end{equation}
where $L=0,1,2,...$ define the multiplet, while $\mu_L$ and $\beta_L$
are production probabilities of mesons and baryons of the given
multiplet in the process of quark hadronization. The
probability $\mu_L$ is determined by characteristic relative momenta
of the fused quarks.

A straightforward way to overcome the ambiguities related to
the resonance production is as follows: one can take into account all
existing resonances and their decay into all possible channels. This is
the scenario suggested in Refs. \cite{10,11}. However, in this way
one faces a set of problems.
The matter is that the number of resonances  which are
observed and cited in the compilation \cite{PDG} is a comparatively
small fraction of the whole set of existing states. The basis for this
statement is provided, for example, by recent investigations of meson
production data
\cite{13,14,15} where a large number of new
meson states with masses in the region 1950--2350 MeV is reported.
One should keep in mind that, naturally, those resonances are first
discovered which can be easily detected.
And one should also recall
that all observed resonances have multiplet partners which are also
produced with approximately equal probabilities and the decay of those
form a background which prevents the direct investigation of Eqs. (1)
and (2).
There exists one more effect which does not allow the realization
of the program discussed in \cite{10,11} at the time being. This is the
effect of the accumulation of widths of overlapping resonances by one
of them; it has been observed for scalar/isoscalar mesons in
the region 1200--1600 MeV \cite{15,17}. As a result of width
accumulation, a broad state $(\Gamma/2 \sim 400$ MeV) has been formed;
similar states can exist, as was stressed in \cite{17}, in other waves,
in other mass regions. Currently it does not seem possible to take into
account the productions and decays of such broad states.

Still, investigation of jet processes in the region of large $x$ opens
a way to test the relations (1) and (2): this article is devoted to a
presentation of this possibility.

The resonance decays increase the
contribution of lighter hadrons, such as pions and kaons in
case of mesons, and nucleons in case of baryons. However, in
jet processes $Z^0 \to q\bar q \to hadrons$ which were
studied in \cite{ALEPH,L3,DELPHI,OPAL},
the spectra, being maximal at $x\sim 0$,
decrease rapidly when $x\to1$. It results in a
rapid increase of contribution of the promptly produced particles
with the increase of $x$, since the decay product carries only a
fraction of $x$ of the initial resonance.
It is just this feature which enables us to estimate the ratio of
probabilities for promptly produced hadrons.

The jets of light hadrons are formed in the processes $Z^0 \to u\bar
u$, $Z^0 \to d\bar d$ and $Z^0 \to s\bar s$. According to \cite{PDG}
\begin{equation}
\Gamma_{u\bar u}/(\Gamma_{u\bar u}+\Gamma_{d\bar d}+\Gamma_{s\bar s})=
0.258\pm 0.031\pm 0.032,
\end{equation}
$$
\Gamma_{d\bar d,s\bar s}/(\Gamma_{u\bar u}+\Gamma_{d\bar
d}+\Gamma_{s\bar s})= 0.371\pm 0.016\pm 0.016.$$

The leading quarks, the probabilities
of which are given by Eq. (4), determine
the hadron content at $x\sim 1$. To find the ratio $V/P$ at large
$x$ we have fitted the spectra $(1/\sigma_{tot})d\sigma/dx$ of
Ref. \cite{ALEPH}
to the sum of exponents $\Sigma C_i e^{-b_ix}$; the results of
the fit
are presented in Fig. 1 for the spectra of $\pi^\pm$, $\pi^0$, $\rho^0$
and $(p,\bar p)$. The ratio of the fited curves, together with
calculation errors, for $\rho^0 /\pi^0$ is shown in Fig. 2a as a shaded
area. We see that in the region $0.6 < x < 0.8$ the data
are in an reasonable agreement with quark combinatorics prediction
$\rho^0/\pi^0=3$.

The same fitting procedure has been done for the $K^{*0}$,
$K^0$, $K^{*\pm}$ and $K^\pm$ spectra.
The figures 2b and 2c demonstrate the ratios $K^{*0}/K^0$ and
$K^{*\pm}/K^\pm$: the data give systematically lower values compared
to the prediction (1), although there is no strong contradiction.

The verification of the barion-to-meson ratio given by Eq. (2) is
of great interest and principal meaning as well. Quark combinatorics
predicts for the proton-to-pion ratio:
\begin{equation}
p/\pi^+\simeq 0.20
\end{equation}
In Fig. 2d one can see
the ratio $p/\pi^+$ given by the fit to the data
\cite{ALEPH} (shaded area) and
the prediction of quark combinatorics (5): the agreement at
$x>0.2$ is quite good.

Let us give a brief comment to the calculation result $p/\pi^+
\simeq 0.20$ for leading particles in jets. In the jet created by a
quark the leading hadrons are produced in the proportions as follows:
\begin{equation}
q_i \to \frac1{4}B_i+\frac12 M_i + \frac1{4}M.
\end{equation}
We consider only the production of hadrons belonging to the lowest
(baryon and meson) multiplets, and, hence, keep only the terms
with $L=0$ in Eq. (3).
In our estimations we assume
$\beta_0\simeq\mu_0$, and therefore we
substitute $B_i\to B_i(0)$, $M_i\to M_i(0)$ and $M\to M(0)$.
The precise content of $B_i(0)$, $M_i(0)$ and $M(0)$ depends on
the proportions in which the sea quarks are produced.
We assume flavour symmetry breaking for sea quarks,
$u\bar u\colon d\bar d\colon s\bar s=1 \colon 1\colon \lambda$,
with  $0\le \lambda \le 1$.
For the sake of simplicity, we put first $\lambda=0$ (actually the ratio
$p/\pi$ depends weakly on $\lambda$). Then for the $u$-quark  initiated
jet we have:
\begin{eqnarray}
&& B_u(0)=\frac2{15}p +\frac1{15}n\ +\
(\Delta-\mbox{resonances }), \\
&& M_u(0)=\frac18\pi^+ +\frac1{16}\pi^0
+\frac1{16}(\eta+\eta') +(\mbox{ vector mesons }),  \nonumber\\
&& M(0)=\frac1{16}\pi^+
+\frac1{16}\pi^0+\frac1{16}\pi^-+\frac1{16} (\eta+\eta')+(\mbox{ vector
mesons }). \nonumber
\end{eqnarray}
The hadron content of the $d$-quark  initiated jet is determined by
isotopic conjugation $p\to n$, $n\to p$, $\pi^+ \to \pi^-$, and  the
content of antiquark jets is governed by charge conjugation; in jets of
strange quarks only sea mesons ($M$) contribute to the ratio
$p/\pi^+$.

Taking into account Eq. (5) and probabilities for the production of
different jets (4), we obtain $p/\pi^+ \simeq 0.21$ for $\lambda=0$.
We can easily get the ratio $p/\pi^+$ for arbitrary
$\lambda$: the decomposition of the ensembles $B_i(0)$, $M_i(0)$,
$M(0)$  with respect to hadron states has been performed in Ref.
\cite{3}, see Appendix D (Tables D.1 and D.2). But, as was stressed
above, this ratio is a weakly  dependent function of $\lambda$:
at $\lambda=1$ we have $p/\pi^+ \simeq 0.20$.

In conclusion, jet-induced processes where the spectra
$(1/\sigma_{tot})d\sigma/dx$ are rapidly decreasing with the growth of
$x$ enable us to perform a model-independent verification of the rules
of quark combinatorics. Experimental data on particle yields for the
decays $Z^0 \to q\bar q \to hadrons$ \cite{ALEPH} are in qualitative
agreement with the predictions of quark combinatorics. Still, the
available experimental data do not allow to conclude
decisively about the accuracy
of the quark combinatorics predictions.

The authors are indebted to V.V. Anisovich for useful discussions.
The paper was partly supported by RFBR grant 98-02-17236.

\newpage

\begin{figure}
\centerline{\epsfig{file=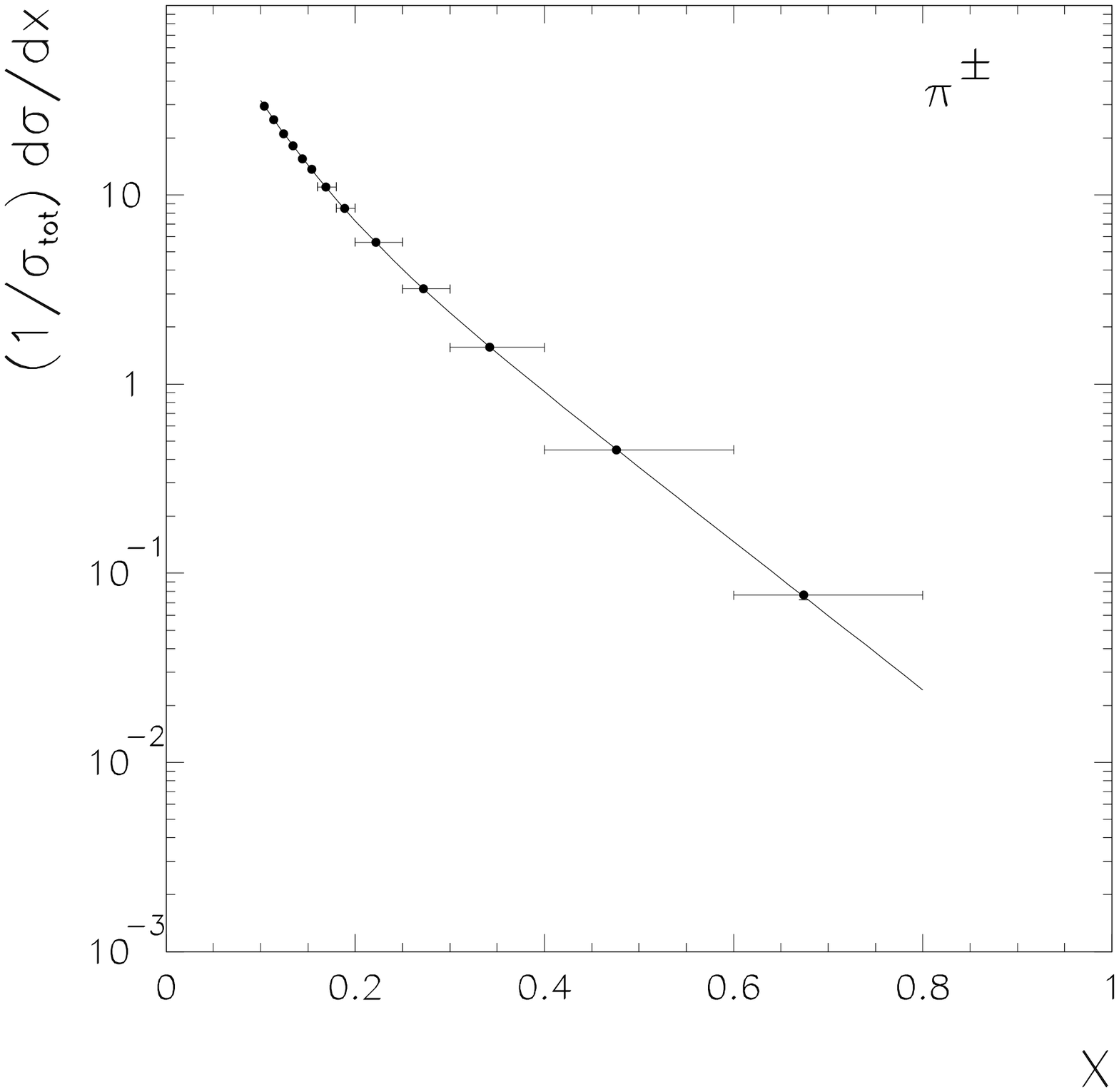 .eps,width=8.0cm}
            \epsfig{file=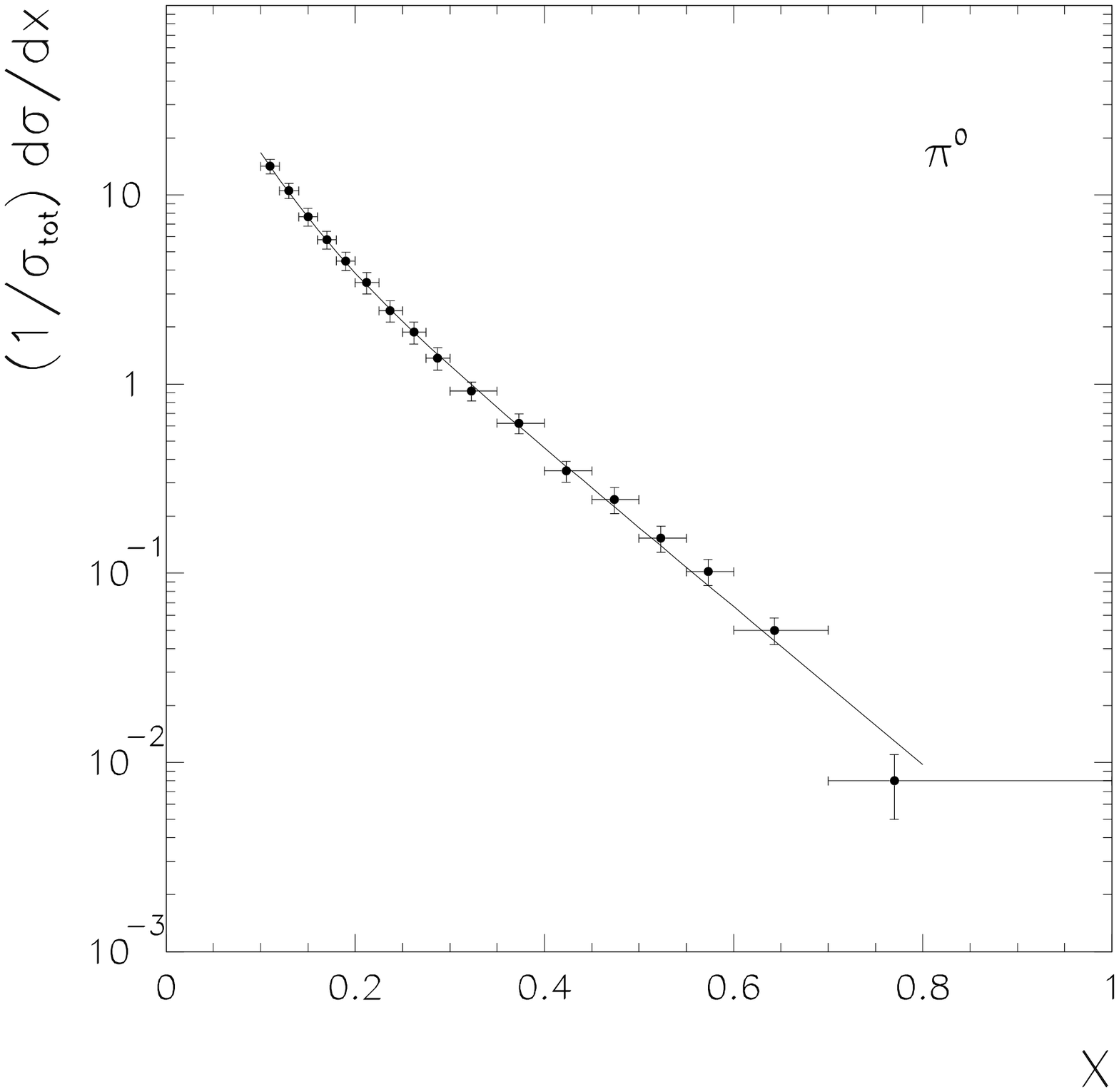  .eps,width=8.0cm}}
\centerline{\epsfig{file=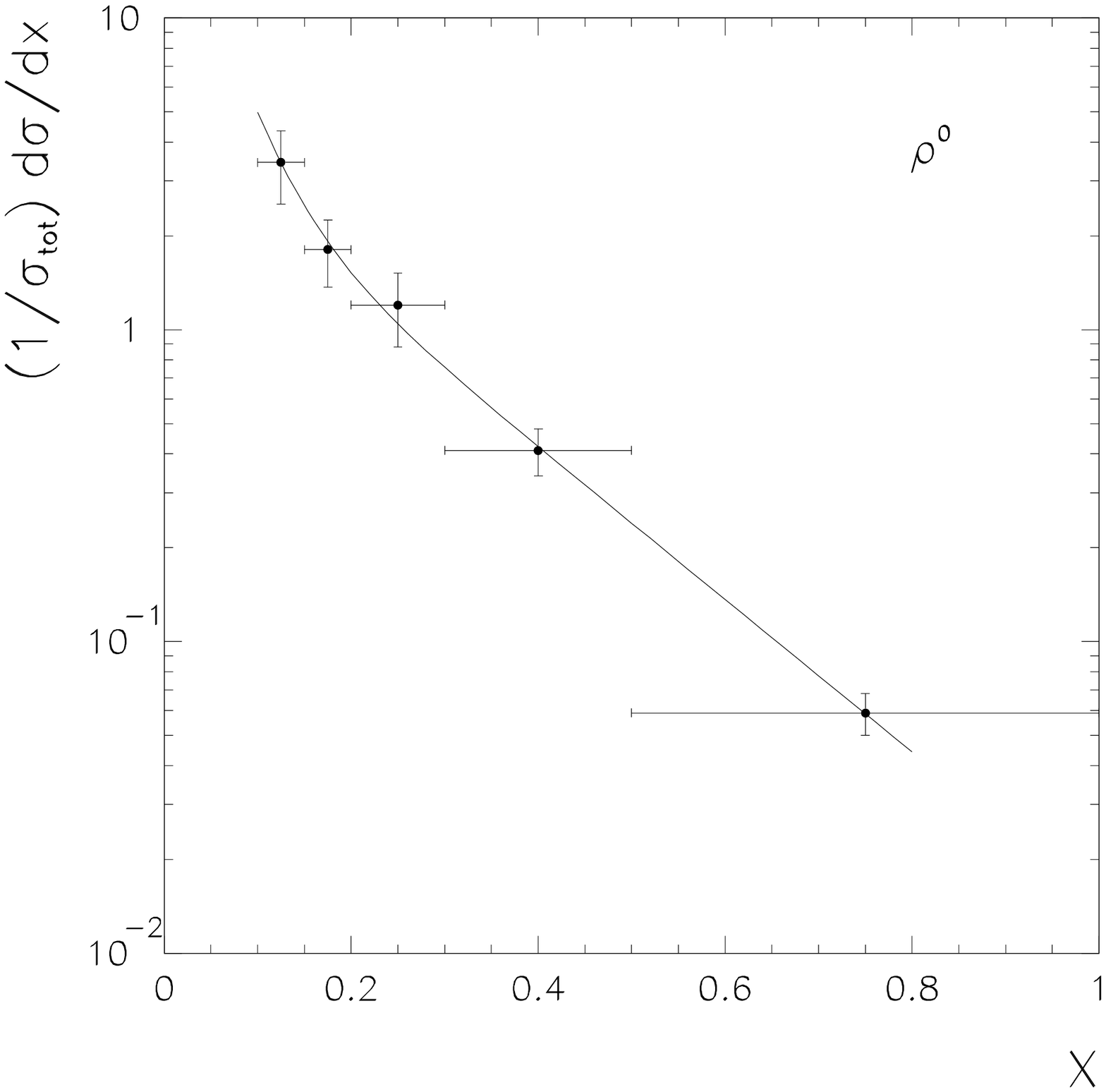 .eps,width=8cm}
            \epsfig{file=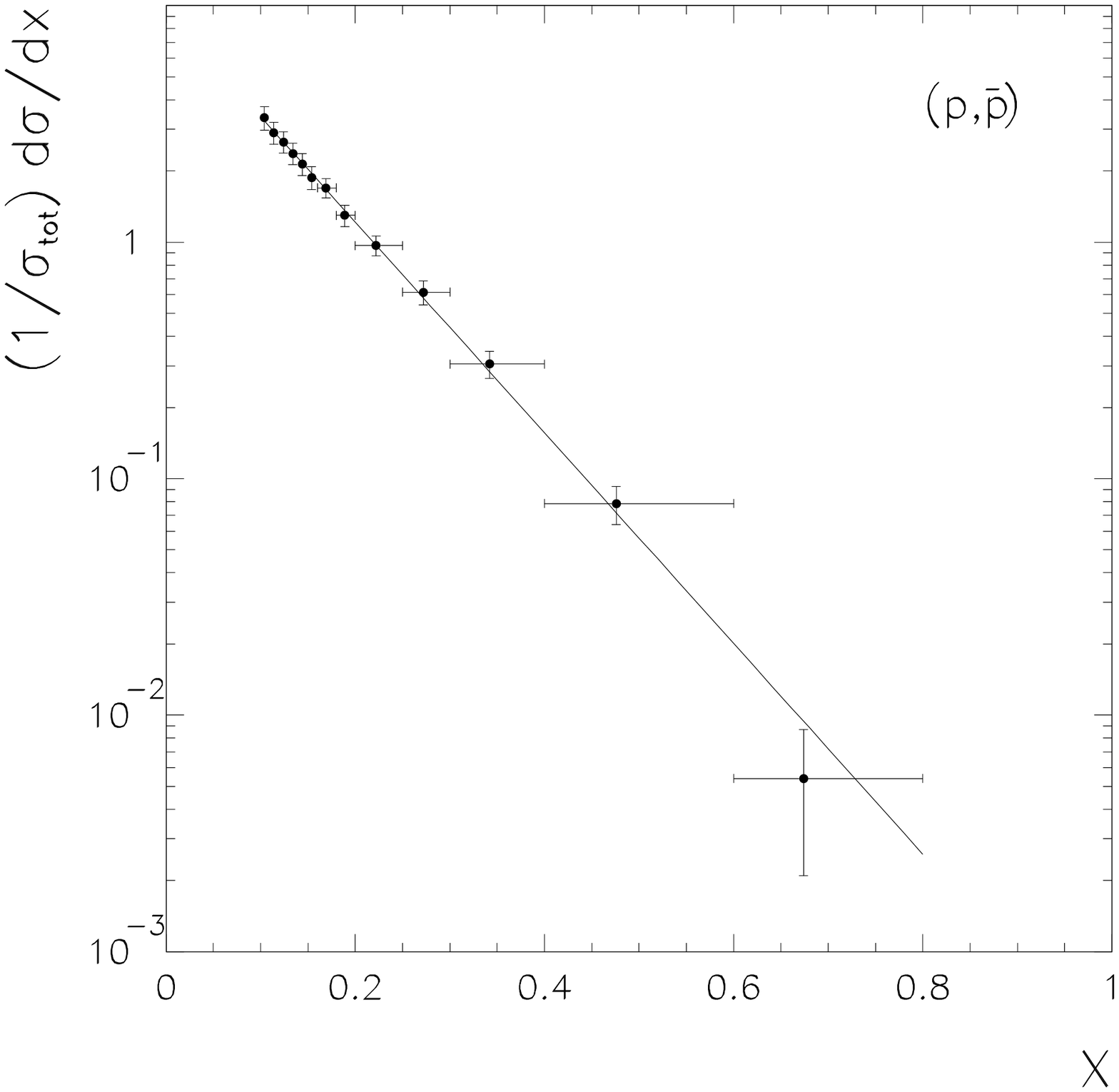,width=8cm}}
\caption{The spectra $(1/\sigma_{tot})\; d\sigma_{tot}/dx$ [16]
for a) $\pi^\pm$, b) $\pi^0$, c) $\rho^0$ and d) $(p,\bar p)$ and their
fit by exponential functions $\Sigma C_i \exp(-b_ix)$.}
\end{figure}

\newpage
\begin{figure}
\centerline{\epsfig{file=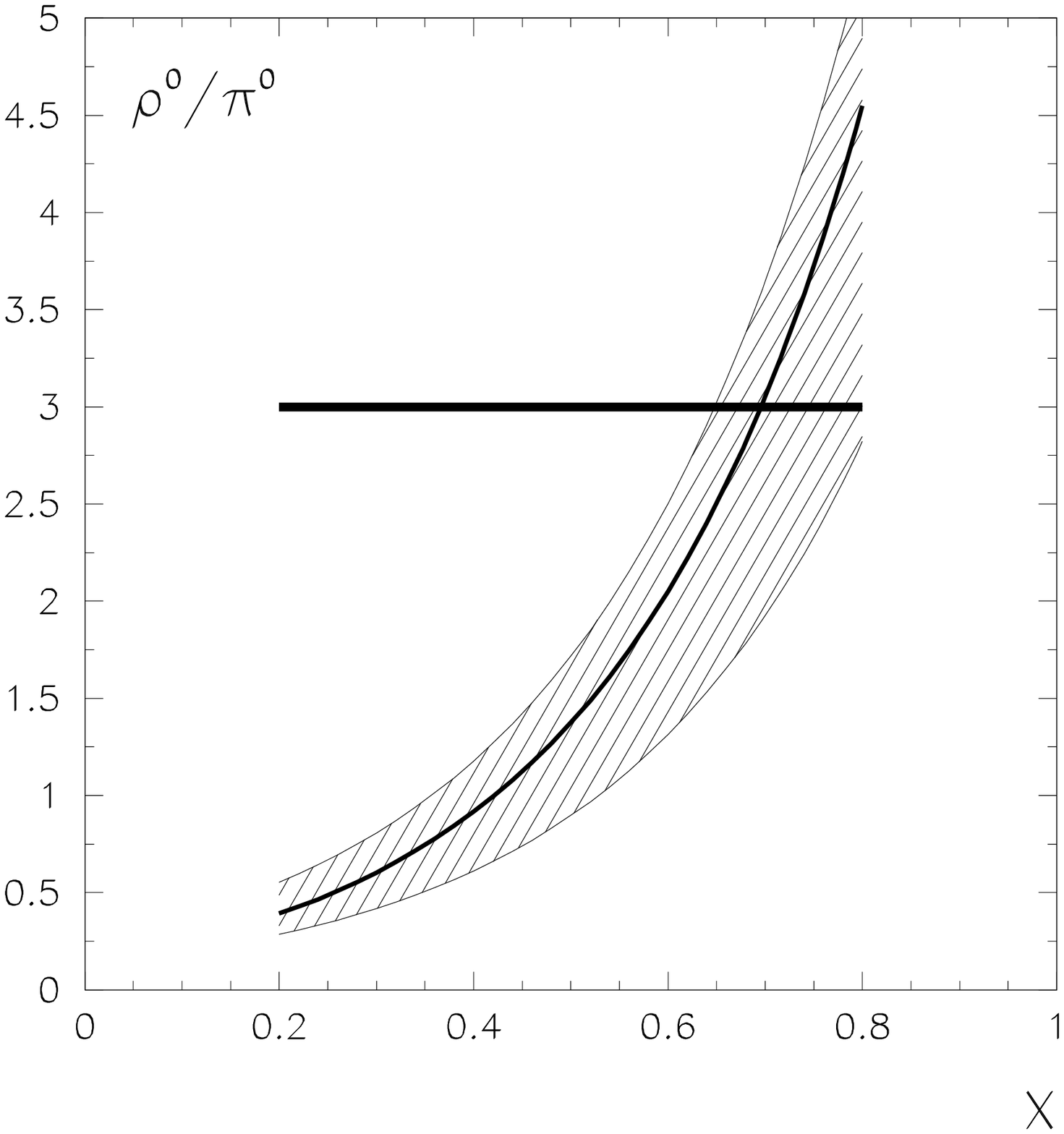,width=8cm}
            \epsfig{file=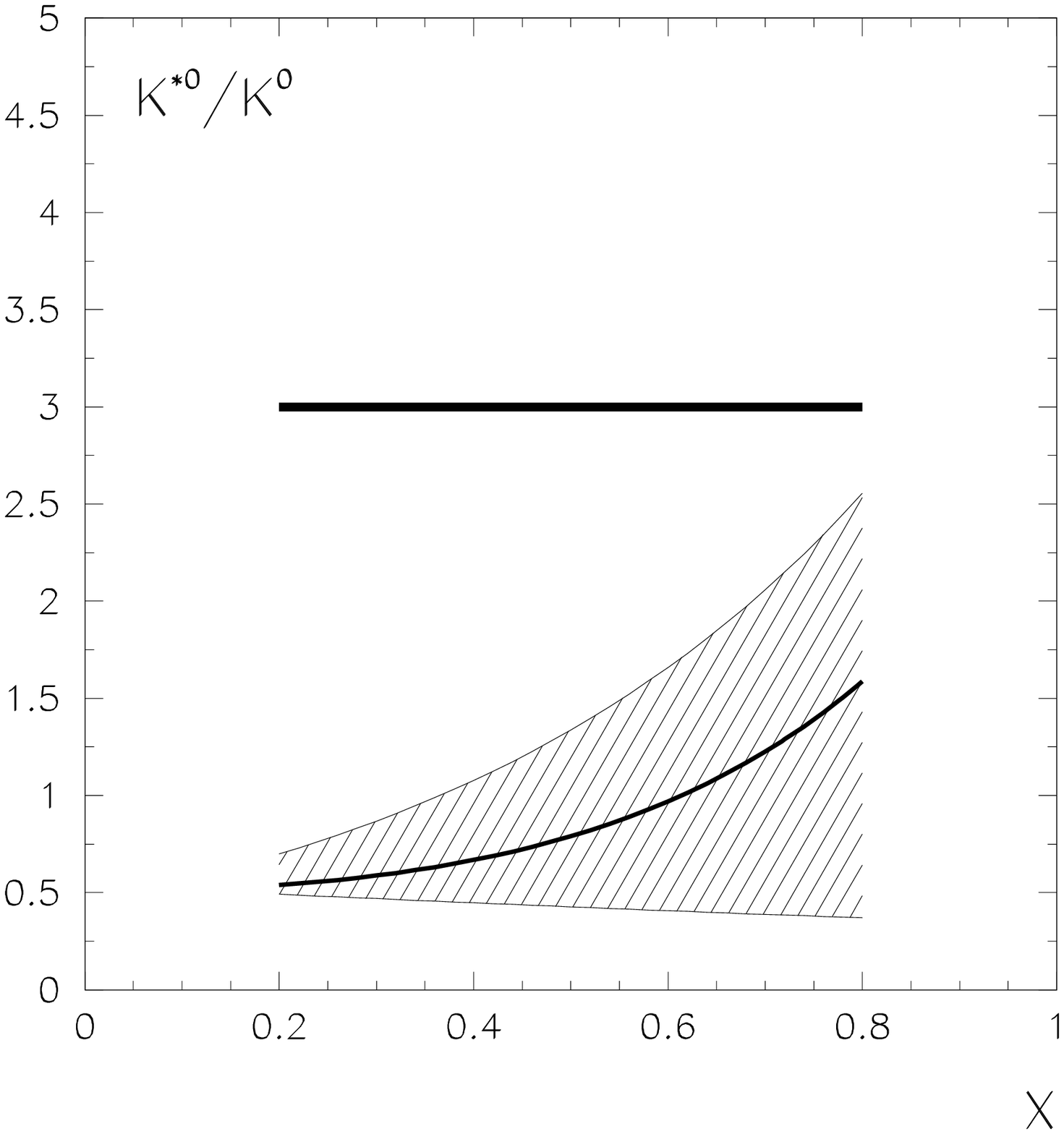,width=8cm}}
\centerline{\epsfig{file=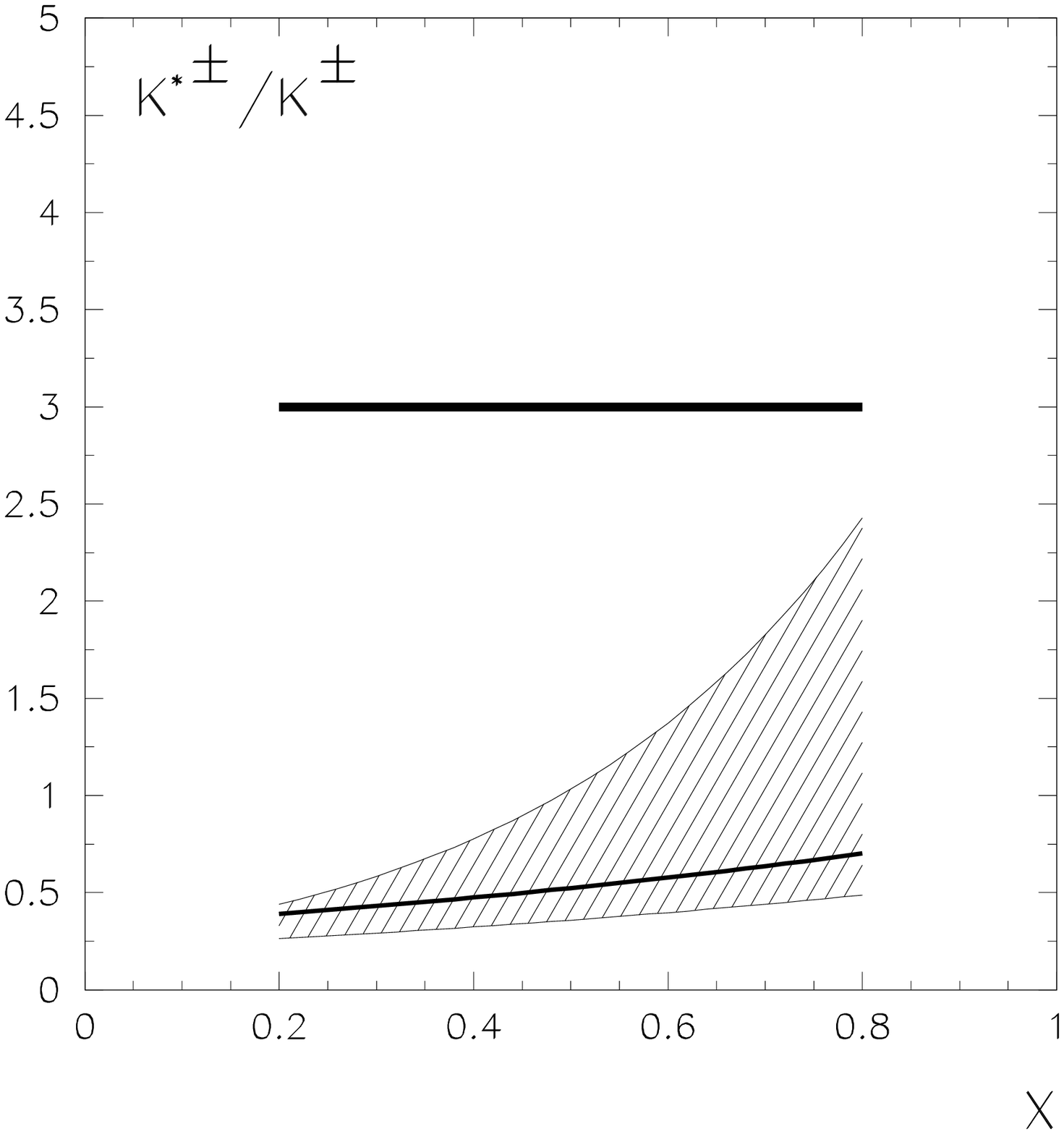,width=8cm}
            \epsfig{file=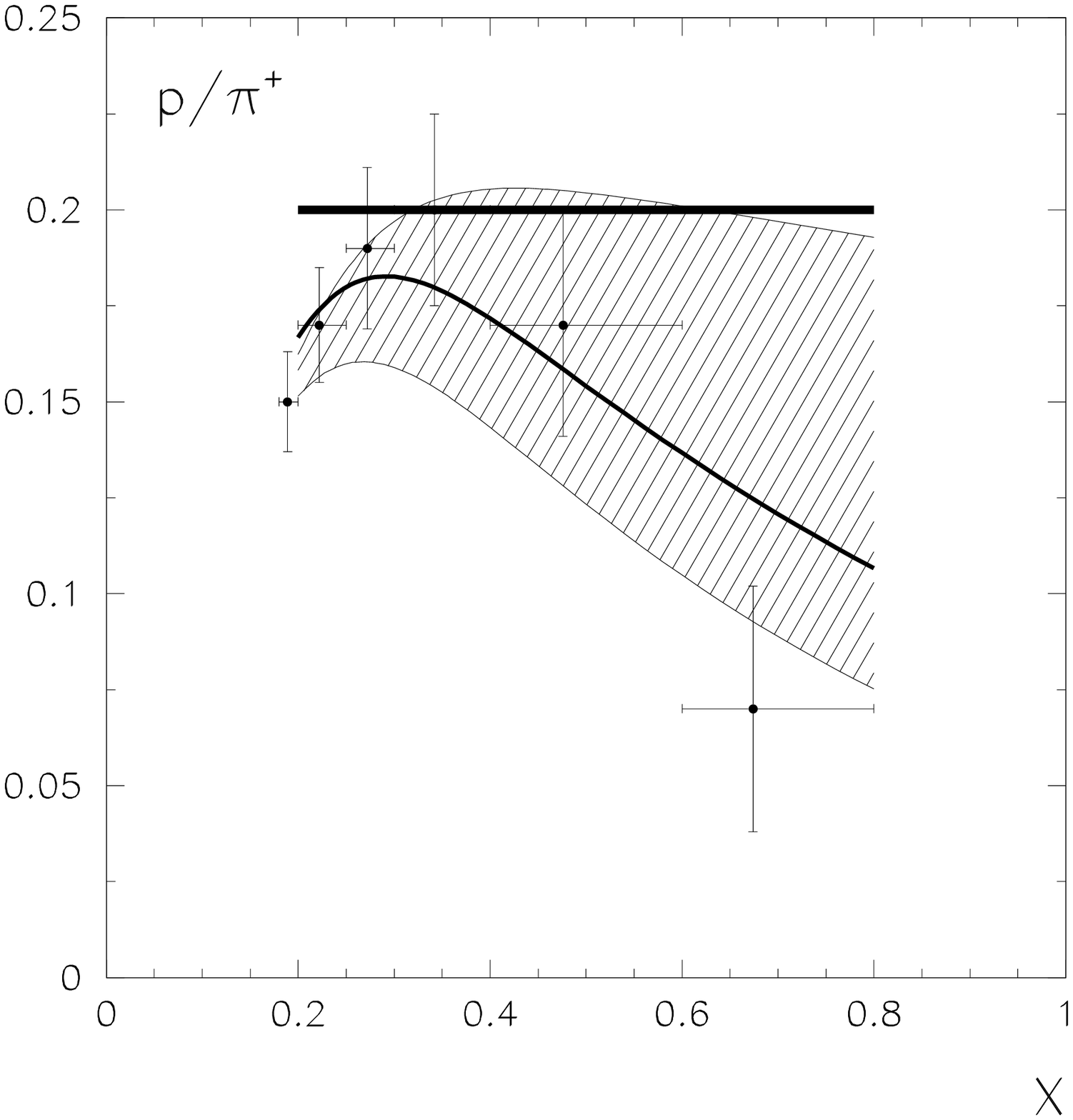 .eps,width=8cm}}
\caption{The ratios $\rho^0/\pi^0$ (a), $K^{*0}/K^0$ (b),
$K^{*\pm}/K^\pm$ (c) and $p/\pi^+$ obtained from the exponetial fit to
ALEPH data [16] (shaded areas). Solid lines stand
for the predictions of quark combinatorics rules. In Fig. 2d the ratio
$p/\pi^+$ is also shown which is obtained using a histogram description
of spectra.}
\end{figure}


\begin{thebibliography}{99}
\bibitem{1}V.V.Anisovich and V.M.Shekhter, Nucl. Phys. {\bf B55}, 455
(1973).
\bibitem{2}J.D. Bjorken and G.E. Farrar, Phys. Rev. {\bf D9}, 1449
(1974).
\bibitem{3} V.V. Anisovich, M.N. Kobrinsky, J. Nyiri and
Yu.M. Shabelski,  "Quark Model and  High Energy Collisions",  World
Scientific, Singapore, 1985.
\bibitem{4}M.A. Voloshin, Yu.P. Nikitin and P.I.Porfirov, Sov. J.
Nucl. Phys. {\bf 35}, 586 (1982).
\bibitem{5}A.K. Likhoded, S.S. Gershtein and Yu.D. Prokoshkin,
Z.  Phys.  {\bf C24}, 305 (1984).
\bibitem{6} C. Amsler and F.E. Close,  Phys. Rev. {\bf D53}, 295 (1996).
\bibitem{7} V.V.Anisovich, Phys. Lett. {\bf B364}, 195 (1995); \\
V.V. Anisovich, Yu.D. Prokoshkin and A.V. Sarantsev,
Phys. Lett. {\bf B382}, 429 (1996); \\
A.V. Anisovich and A.V. Sarantsev, Phys. Lett. {\bf B413}, 137 (1997).
\bibitem{10}Yi-Jin  Pei, Z. Phys. {\bf C72} 39 (1996).
\bibitem{11}P.V. Chliapnikov, Phys. Lett. {\bf B462}, 341 (1999).
\bibitem{PDG} Particle Data Group, C. Caso et al.,
Eur. Phys. J. {\bf C3}, 1 (1998).
\bibitem{13}
C.A. Baker, C.J. Batty, P. Bl\"um  et al., Phys. Lett. {\bf B449}, 114
(1999).
\bibitem{WA} D. Barberis, F.G. Binon, F.E. Close et al., WA
102 Coll., {\it "A study of the $f_0(980)$,  $f_0(1370)$,  $f_0(1500)$,
$f_0(2000)$, and $f_2(1950)$ observed in the centrally produced $4\pi$
states",} hep-ex/0001017 (2000).
\bibitem{14}
A.V. Anisovich, C.A. Baker, C.J. Batty et al., Phys.
Lett. {\bf B452}, 187 (1999); $ibid$, {\bf B452}, 173 (1999);
$ibid$, {\bf B452}, 180 (1999); Nucl. Phys. {\bf A651}, 253 (1999).
\bibitem{15}
A.V.Anisovich, V.V.Anisovich, Yu.D.Prokoshkin and A.V.Sarantsev,
 Z. Phys. {\bf A357}, 123 (1997); \\
A.V.Anisovich, V.V.Anisovich, and A.V.Sarantsev,
Z. Phys. {\bf A359}, 173 (1997); $ibid$,
Phys. Lett.  {\bf B395}, 123 (1997).
\bibitem{17}
V.V. Anisovich, D.V. Bugg, and A.V. Sarantsev,
 Phys. Rev. {\bf D58}:111503  (1998).
\bibitem{ALEPH} ALEPH Coll., R. Barate et al., Phys. Rep. {\bf 294}, 1
(1998).
\bibitem{L3} L3 Coll., M. Acciarri et al., Phys. Lett. {\bf B393}
465 (1997); $ibid$, {\bf B407}, 389 (1997).
\bibitem{DELPHI} DELPHI Coll., P. Abreu et al., Phys. Lett. {\bf B475},
429 (2000); $ibid$,  {\bf B449}, 364 (1999).
\bibitem{OPAL} OPAL Coll., K. Ackerstaff et al., Eur. Phys. J.
{\bf C5}, 411 (1998);  $ibid$, Eur. Phys. J. {\bf C4}, 19 (1998); \\
G. Alexander et al., Z. Phys. {\bf C73}, 569, 587 (1997).
\end{thebibliography}
\end{document}